\documentclass[11pt]{article}
\usepackage{moriond,epsfig}

\usepackage{feynmf}

\def\be{\begin{equation}}
\def\ee{\end{equation}}
\def\bea{\begin{eqnarray}}
\def\eea{\end{eqnarray}}

\newcommand{\text}[1]{\mathrm{#1}}
\newcommand{\hc}[1]{#1^{\dagger}}

\begin{document}
\vspace*{4cm} 
\title{Dark Matter and Electroweak Symmetry Breaking from $SO(10)$}

\author{ K. KANNIKE }

\address{National Institute of Chemical Physics and Biophysics, Ravala pst 10, 10143 Tallinn, Estonia}

\maketitle

\abstracts{
We consider a minimal model of GUT scalar dark matter (DM) stabilized by the discrete gauge matter parity $P_{X}$ that arises from breaking of $SO(10)$. The dark sector comprises the complex singlet $S$ and the inert doublet $H_{2}$. GUT scale parameters are evaluated to the electroweak scale via Renormalization Group Equations (RGEs). Experimental and theoretical constraints limit the DM mass to the 80~GeV to 2~TeV range. The EW symmetry breaking is radiative and can occur via RGE running and 1-loop matching corrections from integrating out DM. Because the next-to-lightest scalar is almost degenerate with DM, it gives a background free displaced decay vertex at the LHC.
}

The Standard Model (SM) is a good theory of ordinary matter. Yet the WMAP measurements of the cosmic microwave background\cite{Komatsu:2008hk} show that $4/5$ of the matter in the Universe is an unknown form of matter -- dark matter (DM) -- usually thought to be a thermal relic whose density is determined by freezeout.

Heavy cold dark matter must be made stable by some symmetry. The simplest such symmetry is a new mirror symmetry or parity $Z(2)$. The usual way to stabilize DM is to impose a global parity by hand. In MSSM, the $R$-parity -- added by hand to prevent fast proton decay \cite{Dimopoulos:1981zb,Farrar:1978xj,Ibanez:1991pr} -- in addition stabilizes neutralino DM. Such $Z(2)$-symmetry is also imposed in low energy phenomenological models of DM with a new scalar singlet \cite{McDonald:1993ex,Burgess:2000yq,Barger:2007im,Barger:2008jx}, doublet \cite{Deshpande:1977rw,Ma:2006km,Barbieri:2006dq,LopezHonorez:2006gr} or higher multiplets \cite{Hambye:2009pw}. 

However, global discrete symmetries are violated by Planck scale operators\cite{Krauss:1988zc}. The solution is to get the $Z_{2}$ from breaking a gauged $U(1)$ embedded in some Grand Unified Theory. One of the most plausible candidates is the $SO(10)$ group that contains the SM symmetry group and an extra $U(1)_{X}$ subgroup. Therefore, $SO(10)$ can broken down to the symmetry group of the Standard Model and the gauged $Z_{2}$ parity
\begin{equation}
P_{X} \equiv P_{M} = (-1)^{3 (B-L)}
\label{eq:PM}
\end{equation} 
that is equivalent to the $R$-parity in supersymmetric theories.

Each generation of SM fermions and the heavy singlet neutrinos needed for the seesaw mechanism of neutrino mass\cite{Gell-Mann:1979kx,Glashow:1979nm,Minkowski:1977sc,Mohapatra:1979ia,Yanagida:1979uq} reside in the representation $\bf 16$ of $SO(10)$. They are odd under the $P_{M}$ parity Eq.~(\ref{eq:PM}). The Standard Model Higgs in $\bf 10$ is even. To be stable, scalar dark matter has to be odd\cite{Kadastik:2009dj} under $P_{M}$. Because the only small representation that is odd under $P_{M}$ is the $\bf 16$, the minimal model of scalar $SO(10)$ DM adds one scalar $\bf 16$ to the theory.

The $SO(10)$ symmetric scalar potential of one $\mathbf{16}$ and one $\bf 10$ is
\begin{equation} 
\begin{array}{rcl}
    V &=& \mu_{1}^{2} \; {\bf 10}\; {\bf 10}
    +  \lambda_{1} ({\bf 10} \; {\bf 10})^2
    + \mu_{2}^{2} \; \overline{{\bf 16}} \; {\bf 16}
    + \lambda_{2} ( \overline{\bf 16}\, {\bf 16})^2 \\
    &+& \lambda_{3}  ({\bf 10} \; {\bf 10}) (\overline{\bf 16} \, {\bf 16} )
    + \lambda_{4} ( {\bf 16\;} {\bf 10}) (\overline{\bf 16}\, {\bf 10} ) \\
    &+& \frac{1}{2} \left( \lambda'_{S}  {\bf 16}^4 + \mathrm{h.c.} \right)
    + \frac{1}{2}  \left( \mu'_{SH} {\bf 16\;} {\bf 10\;} {\bf 16}
    + \mathrm{h.c.} \right).
\end{array}
\label{eq:V:GUT}
\end{equation}
All the parameters are real with the exception of $\lambda'_{S}$ and $\mu'_{SH}$. We assume that $SO(10)$ breaks down to $SU(3)_{c} \times SU(2)_{L} \times U(1)_{Y} \times P_{M}$ in such a way that only one SM Higgs boson $H_{1} \in \mathbf{10}$ and the DM candidates complex singlet $S \in \mathbf{16}$ and the Inert Doublet $H_{2} \in \mathbf{16}$ are light, but all other particles have masses of order $M_\mathrm{G}$.

Below $M_\mathrm{G}$, the most general CP-invariant scalar potential invariant under the $P_{M}$ parity $H_{1} \to H_{1}$, $H_{2} \to -H_{2}$, $S \to -S$ is
\begin{equation}
\begin{array}{rcl}
    V &=& \mu_{1}^{2} \hc{H_{1}} H_{1} + \lambda_{1} (\hc{H_{1}} H_{1})^{2}
    + \mu_{2}^{2} \hc{H_{2}} H_{2} + \lambda_{2} (\hc{H_{2}} H_{2})^{2} +
    \mu_{S}^{2} \hc{S} S 
    \\
    &+&  \frac{\mu_{S}^{\prime 2}}{2} \left[ S^{2} + (\hc{S})^{2} \right] +
    \lambda_{S} (\hc{S} S)^{2} + \frac{ \lambda'_{S} }{2} \left[ S^{4} + (\hc{S})^{4} \right]
    + \frac{ \lambda''_{S} }{2} (\hc{S} S) \left[ S^{2} + (\hc{S})^{2} \right] \\
    &+& \lambda_{S1}( \hc{S} S) (\hc{H_{1}} H_{1}) + \lambda_{S2} (\hc{S} S) (\hc{H_{2}} H_{2}) \\
    &+& \frac{ \lambda'_{S1} }{2} (\hc{H_{1}} H_{1}) \left[ S^{2} + (\hc{S})^{2} \right]
    + \frac{ \lambda'_{S2} }{2} (\hc{H_{2}} H_{2}) \left[ S^{2} + (\hc{S})^{2} \right] \\
    &+& \lambda_{3} (\hc{H_{1}} H_{1}) (\hc{H_{2}} H_{2})
    + \lambda_{4} (\hc{H_{1}} H_{2}) (\hc{H_{2}} H_{1})
    + \frac{\lambda_{5}}{2} \left[(\hc{H_{1}} H_{2})^{2} + (\hc{H_{2}} H_{1})^{2} \right] \\
    &+& \frac{\mu_{S H}}{2}  \left[\hc{S} \hc{H_{1}} H_{2} + \hc{H_{2}} H_{1} S \right]
    + \frac{\mu'_{S H}}{2}  \left[S \hc{H_{1}} H_{2} + \hc{H_{2}} H_{1} \hc{S} \right],
\end{array}
\label{eq:V:CP:inv}
\end{equation}
together with the GUT scale boundary conditions
\begin{eqnarray}
   \mu_1^2(M_{\mathrm{G}}) &>& 0,\; \mu_2^2(M_{\mathrm{G}})
    =\mu_S^2(M_{\mathrm{G}}) >0, \nonumber \\
    \lambda_2(M_{\mathrm{G}}) &=& \lambda_S(M_{\mathrm{G}})
    =\lambda_{S2}(M_{\mathrm{G}}),
    \; \lambda_3(M_{\mathrm{G}})=\lambda_{S1}(M_{\mathrm{G}}),
    \label{eq:gut:parameters:equal}
\end{eqnarray}
and
\begin{eqnarray}
    \mu_{S}^{\prime 2}, \mu_{SH}^{2}
    &\leq& {{\mathcal O} \frac{M_{\mathrm{G}}}{M_\mathrm{P}} }^{n} \mu^{2}_{1,2}, \nonumber \\
    \lambda_{5}, \lambda'_{S1} , \lambda'_{S2} , \lambda''_{S}
    &\leq& {{\mathcal O} \frac{M_\mathrm{G}}{M_\mathrm{P}} }^n \lambda_{1,2,3,4}.
    \label{eq:gut:parameters:zero}
\end{eqnarray}
The parameters in Eq.~(\ref{eq:gut:parameters:zero}) can only be generated by operators suppressed by $n$ powers of the Planck scale $M_\mathrm{P}$.

\begin{figure}[htbp]
\begin{center}
\begin{fmffile}{diagrams}
\fmfframe(3,1)(4,0){
\begin{fmfgraph*}(20,15)
\fmfleft{i1}
\fmfright{o1}
\fmflabel{$H_{1}$}{i1}
\fmflabel{$H_{1}$}{o1}
\fmf{scalar,tension=2}{i1,v1}
\fmf{scalar,tension=2}{v2,o1}
\fmf{fermion,label=$t$,label.side=left,left}{v1,v2}
\fmf{fermion,label=$t$,label.side=left,left}{v2,v1}
\end{fmfgraph*}
}
\fmfframe(4,1)(0,0){
\begin{fmfgraph*}(20,15)
\fmfleft{i1}
\fmfright{o1}
\fmflabel{$H_{1}$}{i1} 
\fmflabel{$H_{1}$}{o1}
\fmf{scalar,tension=2}{i1,v1}
\fmf{scalar,tension=2}{v2,o1}
\fmf{scalar,label=$S$,label.side=left,left}{v1,v2}
\fmf{scalar,label=$H_{2}$,label.side=right,right}{v1,v2}
\end{fmfgraph*}
}
\end{fmffile}
\caption{Dominant diagrams contributing to the Higgs boson mass.}
\label{fig:diagrams}
\end{center}
\end{figure}
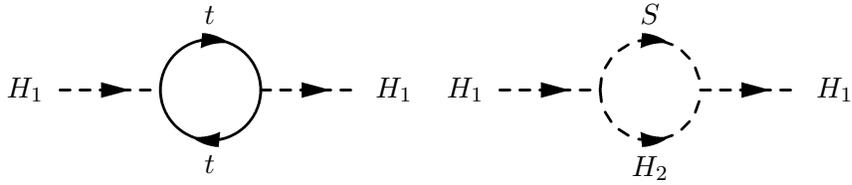

We see that the dimensionful coupling $\mu'_{SH}$, not suppressed by $SO(10)$, can be large and form a ``soft portal'' to the dark sector \cite{Patt:2006fw}. It can induce electroweak symmetry breaking\cite{Kadastik:2009ca} via the diagrams in Fig.~\ref{fig:diagrams}. (EWSB via effective potential for the Inert Doublet Model was considered in \cite{Hambye:2007vf}.)

In the mass spectrum of the dark sector we have the charged Higgs from $H_{2}$ and four neutral mass eigenstates from the mixing of the singlet and doublet neutral components.
The mass matrices of real and imaginary neutral scalars, respectively, are
\begin{equation}
    M_{H,A}^{2} =
\left(
	\begin{array}{cc}
    \mu_{S}^{2} \pm \mu_{S}^{\prime 2} + \lambda_{S1} v^2/2
    & \pm \mu'_{SH} v/(2 \sqrt{2}) \\
     \pm \mu'_{SH} v/(2 \sqrt{2})
    & \mu_{2}^{2}+ (\lambda_{3} + \lambda_{4}) v^{2}/2
    \end{array}
   \right),
\label{eq:M:S:real:pot}
\end{equation}
where we have neglected all $SO(10)$-suppressed parameters save $\mu_{S}^{\prime 2}$. The mass spectrum is $M_{\mathrm{DM}} \leq M_{\mathrm{NL}} \ll M_{\mathrm{NL2}} \leq M_{\mathrm{NL3}}$, where the next-to-lightest (NL) particle is almost degenerate with DM. There is another, heavier, pair of states $S_{\mathrm{NL2}}$ and $S_{\mathrm{NL3}}$. The mass gaps between $M_{\mathrm{DM}}, M_{\mathrm{NL}}$ and likewise between $M_{\mathrm{NL2}}, M_{\mathrm{NL3}}$ are proportional to $\mu_{S}^{\prime 2}$. We give $\mu_{S}^{\prime 2}$ a small positive value to avoid total degeneracy of real and imaginary mass eigenstates.

Because we have a lot of unknown parameters, we do a Monte Carlo scan over them at the GUT scale and run them down to the electroweak scale by renormalization group equations. We integrate out the dark sector particles at their average mass and the top quark at its mass scale.

At the GUT scale we impose $SO(10)$ boundary conditions (\ref{eq:gut:parameters:equal}) and (\ref{eq:gut:parameters:zero}). In addition we demand that the electroweak symmetry breaking must arise from dark matter. We require perturbativity of dimensionless interactions ($\lambda_{i} < 4 \pi$) and vacuum stability in the whole range from $M_{Z}$ to $M_{\mathrm{GUT}}$. LEP2 data says that $H^{+}$ must be heavier than about 80~GeV, and we have a lower bound of $M_{Z}/2$ on dark matter mass from $Z$ invisible width\cite{Amsler:2008zzb}. Last not least, dark matter must have correct cosmic density within $3\sigma$, that is $0.94 < \Omega_{\mathrm{DM}} < 0.129$\cite{Komatsu:2008hk}.

\begin{figure}[htbp]
\begin{center}
\epsfig{file=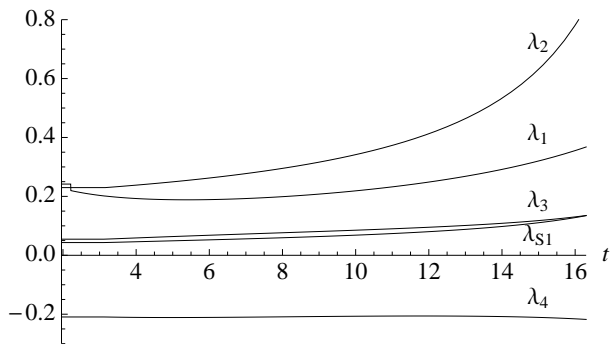,width=0.5\textwidth}
\caption{An example of running interaction couplings.}
\label{fig:lambda:run}
\end{center}
\end{figure}

Fig.~\ref{fig:lambda:run} shows an example of running of the dimensionless interaction couplings from GUT scale to the electroweak scale. 

\begin{figure}[htbp]
\begin{center}
\epsfig{file=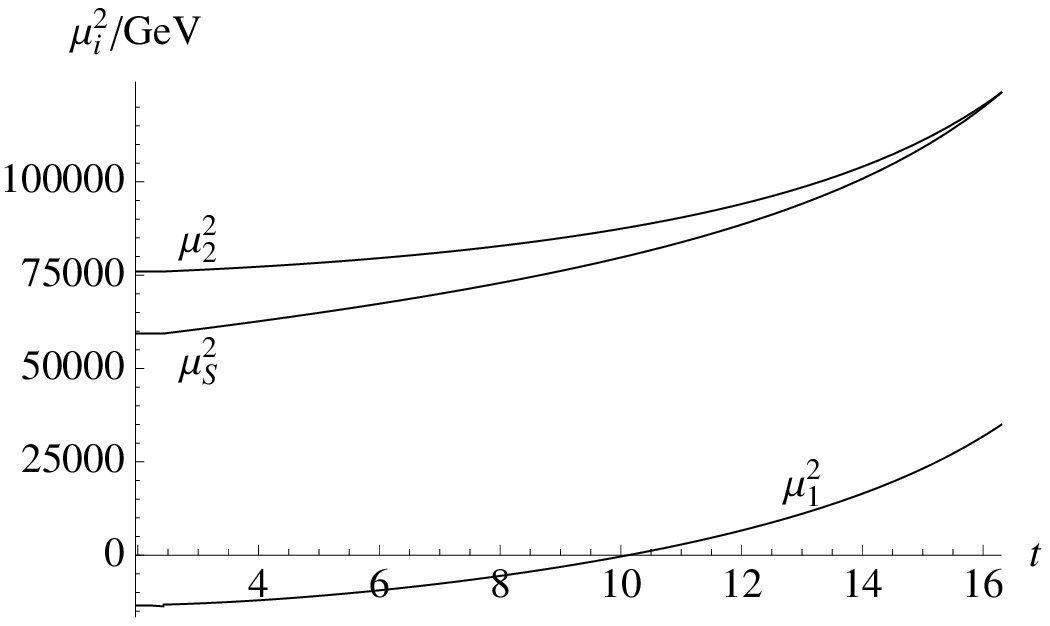,width=0.48\textwidth}
\epsfig{file=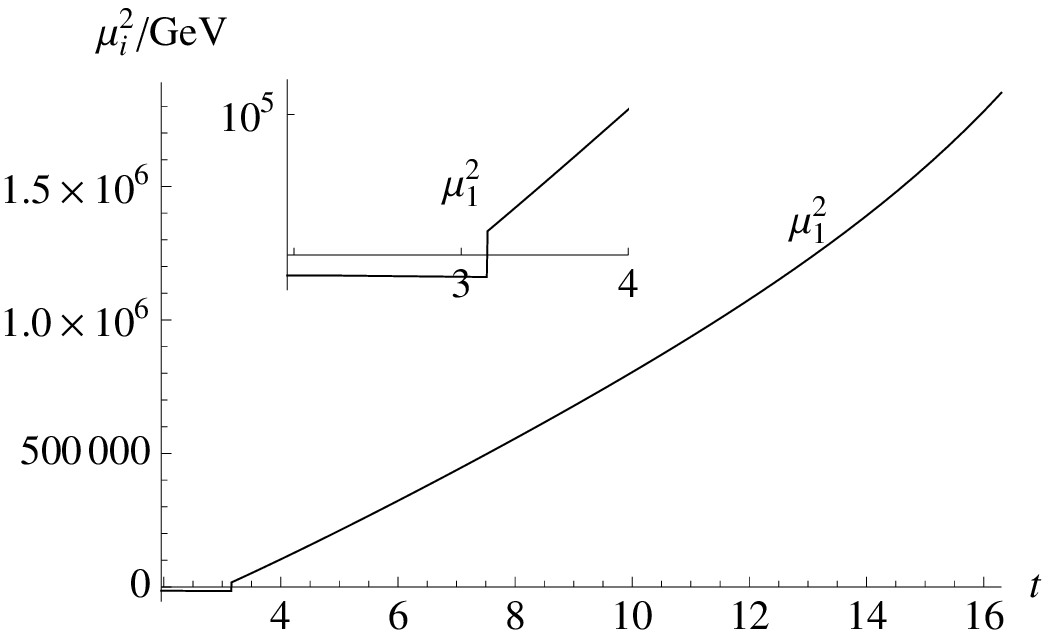,width=0.48\textwidth}
\caption{Examples of running mass parameters. On the left panel, EWSB is achieved via RGE running, on the right panel, via 1-loop corrections from DM, as shown on the inset.}
\label{fig:mu:run}
\end{center}
\end{figure}

Fig.~\ref{fig:mu:run} displays running of mass parameters. Note that we have two distinct possibilities to induce electroweak symmetry breaking: (i) via the Higgs mass parameter $\mu_{1}^{2}$ evolving to negative values via RGE running, and (ii) by integrating out dark matter in the effective potential\cite{Coleman:1973jx,Casas:1998cf}, equivalent to calculating the 1-loop diagrams shown in Fig.~\ref{fig:diagrams}. The first possibility is demonstrated on the left panel and the second one on the right panel of Fig.~\ref{fig:mu:run}. The loop mechanism is embedded in the $SO(10)$ GUT context here, but it is a general mechanism that can as well originate in some low energy effective theory.

\begin{figure}[htbp]
\begin{center}
\epsfig{file=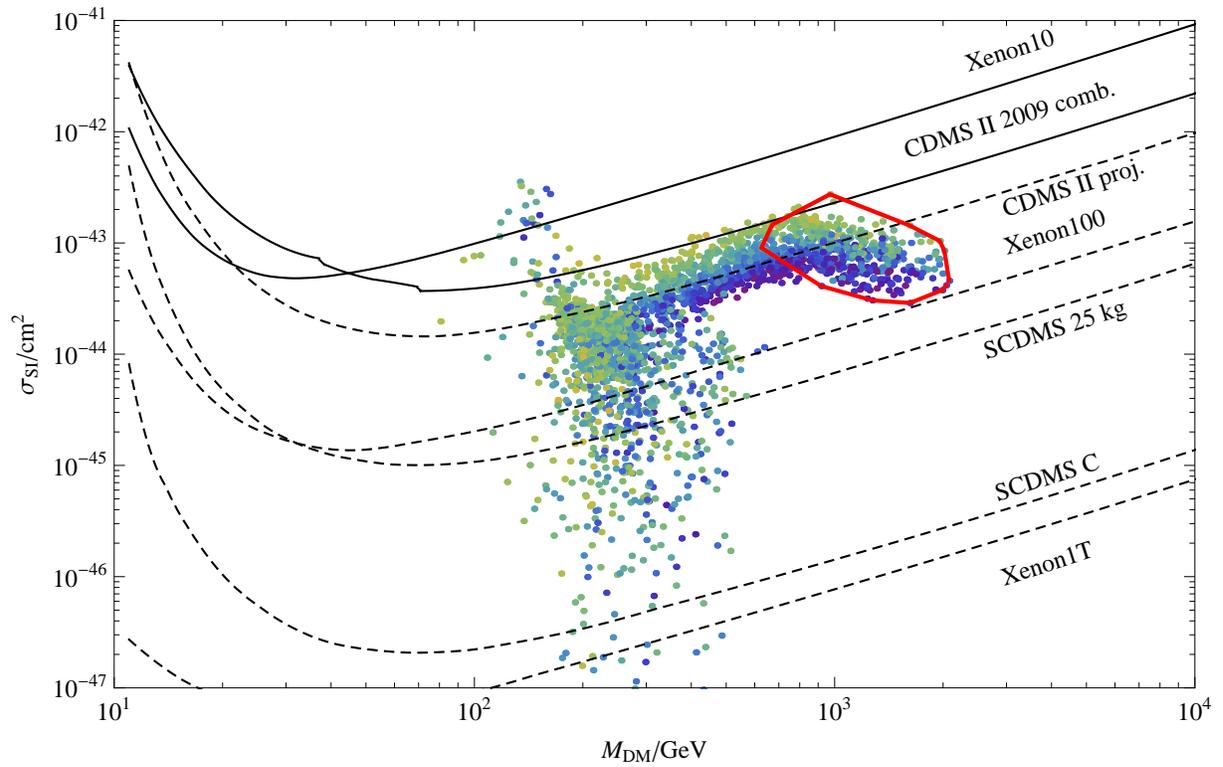,width=\textwidth}
\caption{DM direct detection cross section per nucleon \textit{vs.} DM mass. The colour signifies Higgs mass from 130~GeV (yellow) to 185~GeV (violet). The region circled by red line allows EWSB by integrating out DM.}
\label{fig:sigma:mdm}
\end{center}
\end{figure}

The Monte Carlo points that satisfy all constraints are plotted on Fig.~\ref{fig:sigma:mdm} as DM mass \textit{vs.} its spin-independent direct detection cross section per nucleon. The solid lines show sensitivities of current experiments like CDMS \cite{Ahmed:2009zw} and Xenon\cite{Angle:2007uj,Angle:2009xb}, the dashed lines are the expected sensitivities of future experiments. In the low mass region, the cross section can vary a lot, because there are several different annihilation reactions, and accidental cancellations can occur in the effective Higgs-DM-DM coupling
\begin{equation}
\lambda_{\text{eff}} \, v = \frac{1}{2} (-\sqrt{2} s\, c\, \mu'_{SH} + 2 s^{2} (\lambda_{3}+\lambda_{4}) v +2 c^{2} \lambda_{S1} v).
\label{eq:lambda:eff:v}
\end{equation}
If dark matter has a relatively high mass, both the annihilation and direct detection cross sections are dominated by the dimensionful Higgs-DM-DM coupling $\mu'_{SH}$. The high mass region in which electroweak symmetry breaking can occur by integrating out dark matter is circled in red.

\begin{figure}[htbp]
\begin{center}
\epsfig{file=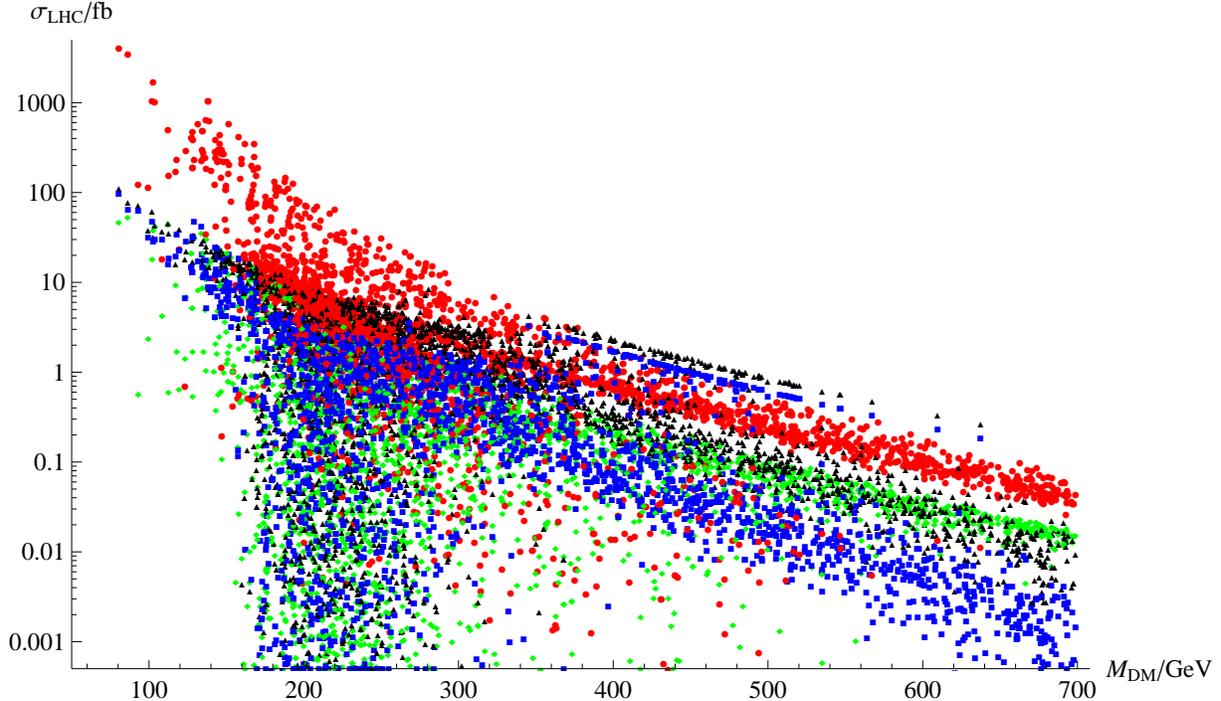,width=\textwidth}
\caption{Dark sector particle production cross sections for LHC with centre of mass energy $\sqrt{s} = \mathrm{14~TeV}$. Red dots mean the process $pp \to S_{\mathrm{NL}} S_{\mathrm{NL}}$, green lozenges $pp \to S_{\mathrm{NL}} S_{\mathrm{NL3}}$, blue squares $pp \to S_{\mathrm{DM}} S_{\mathrm{NL}}$, and black triangles $pp \to S_{\mathrm{NL}} H^{\pm}$.}
\label{fig:dm:prod:lhc}
\end{center}
\end{figure}

If the dark sector particles are relatively light (with masses up to about 700~GeV), they can be produced in the LHC\cite{Kadastik:2009gx}. The Fig.~\ref{fig:dm:prod:lhc} shows LHC production cross sections for the  processes $pp \to S_{\mathrm{NL}} S_{\mathrm{NL}}$, $pp \to S_{\mathrm{NL}} S_{\mathrm{NL3}}$,  $pp \to S_{\mathrm{DM}} S_{\mathrm{NL}}$, and $pp \to S_{\mathrm{NL}} H^{\pm}$. The first three reactions generate dark sector particles from quark-quark interactions (via $Z^{*}$) or gluon fusion (via $h^{*}$), the last one from quarks via $W^{\pm *}$. The cross sections are correlated with direct detection cross sections, so if dark matter is discovered in CDMS II or Xenon100, we can hope it can be detected at the LHC as well.

Because the mass splitting between dark matter and the next-to-lightest particle is suppressed by $SO(10)$, the next-to-lightest particle can have a long lifetime and give a vertex displaced from the collision point by millimetres to metres, decaying into dark matter and a pair of leptons. This is a highly distinct signature that is easy to discover.

In conclusion, we consider breaking non-SUSY $SO(10)$ GUT into the SM symmetry group and the matter parity $P_{M}$. The new parity is not a global symmetry imposed by hand but a discrete gauge symmetry. The dark matter resides in a scalar representation $\bf 16$ of $SO(10)$. Because it is odd under $P_{M}$, it is the scalar analogue of Standard Model fermions. We require DM to induce electroweak symmetry breaking. This and other constraints predict a DM mass range between 80~GeV and 2~TeV. The collider signature of the dark sector is a displaced vertex of two leptons with almost no background.

\section*{Acknowledgments}

This work was supported by the ESF Grant 8090, JD164, SF0690030s09 and EU FP7-INFRA-2007-1.2.3 contract No 223807.

\end{document}